\documentclass[review]{elsarticle}

\usepackage{lineno,hyperref}
\usepackage{algorithm}
\usepackage{algpseudocode}
\usepackage{amsmath}
\usepackage{amssymb}
\modulolinenumbers[5]

\journal{Computers \& Geosciences}

\begin{document}

\begin{frontmatter}

\title{Multichannel Analysis of Surface Waves Accelerated (MASWAccelerated): Software for Efficient Surface Wave Inversion Using MPI and GPUs}

\author{Joseph Kump*, Eileen R. Martin\footnote{Authorship Statement: Joseph Kump was the primary software and algorithm developer and contributed to writing this manuscript. Eileen Martin determined the initial problem statement, advised on software development and testing and contributed to writing this manuscript. }}
\address{Virginia Polytechnic Institute and State University \\ Department of Mathematics, McBryde Hall, 225 Stanger St., Blacksburg, VA 24060 \\ *Corresponding author: \url{josek97@vt.edu} \\ Code available at \url{https://github.com/jlk9/MASWA}}

\begin{abstract}
Multichannel Analysis of Surface Waves (MASW) is a technique frequently used in geotechnical engineering and engineering geophysics to infer layered models of seismic shear wave velocities in the top tens to hundreds of meters of the subsurface. We aim to accelerate MASW calculations by capitalizing on modern computer hardware available in the workstations of most engineers: multiple cores and graphics processing units (GPUs). We propose new parallel and GPU accelerated algorithms for evaluating MASW data, and provide software implementations in C using Message Passing Interface (MPI) and CUDA. These algorithms take advantage of sparsity that arises in the problem, and the work balance between processes considers typical data trends. We compare our methods to an existing open source Matlab MASW tool. Our serial C implementation achieves a 2x speedup over the Matlab software, and we continue to see improvements by parallelizing the problem with MPI. We see nearly perfect strong and weak scaling for uniform data, and improve strong scaling for realistic data by repartitioning the problem to process mapping. By utilizing GPUs available on most modern workstations, we observe an additional 1.3x speedup over the serial C implementation on the first use of the method. We typically repeatedly evaluate theoretical dispersion curves as part of an optimization procedure, and on the GPU the kernel can be cached for faster reuse on later runs. We observe a 3.2x speedup on the cached GPU runs compared to the serial C runs. This work is the first open-source parallel or GPU-accelerated software tool for MASW imaging, and should enable geotechnical engineers to fully utilize all computer hardware at their disposal. 
\end{abstract}

\begin{keyword}
Surface Waves \sep Near-surface Imaging \sep GPU \sep Parallel Computing \sep MPI
\end{keyword}


\end{frontmatter}



\section{Introduction and Background}
Multichannel Analysis of Surface Waves (MASW) is a seismic exploration technique used to infer a layered 1D model of the subsurface. In MASW a seismic source is recorded by a linear array of geophones or other vibration sensors, then a dispersion curve is calculated from that data, indicating that Rayleigh waves at each of the wavelengths of interest, $W$, travel at phase velocities, $C_e$ \cite{Park1999}, \cite{Louie2001}. To infer whether any test velocity model of the subsurface would explain the observed surface wave dispersion, $C_e$, a theoretical dispersion curve, $C_t$ is modeled based on how surface waves would be expected to propagate in that test velocity model. This procedure is repeated many times in an optimization loop, minimizing the difference between $C_e$ and $C_t$, the current test velocity's theoretical dispersion curve. This technique is widely used in geotechnical engineering for site investigation, particularly because it does not require drilling cores or samples, permitting is simpler, and it can be applied to both active controlled-source and passively recorded seismic signals \cite{Park2007}. 

There exists open-source code available to geotechnical engineers, for example, MASWaves which is implemented in Matlab \cite{Olafsdottir2018}. However, to our knowledge there is not currently any open-source MASW software that specifically focuses on optimizing performance. Thus we are motivated to implement MASW in C with a focus on fast performance. Faster analysis of MASW data means that a single theoretical dispersion curve could be evaluated faster, but more importantly it would reduce the cost of repeated evaluations of theoretical dispersion curves. We always perform MASW modeling as part of a larger optimization problem to find the ``best" velocity model to explain our data. Further, engineers are interested in quantifying the uncertainty in subsurface velocity models. Both scenarios require repeated evaluations of MASW modeling, so any inefficiencies in existing codes repeatedly accumulate into significant additional electricity cost and wait time for engineers.

As we expand our problem sizes to larger datasets (potentially having finer-scale dispersion curves, or extending over wider frequency ranges), we can take advantage of computer clusters which have many cores to drastically reduce the time to actionable solutions. A cluster environment is already the current scenario for rapidly evaluating uncertainty in dispersion images \cite{dou2017}. Further, the majority of laptops today have multiple cores, so parallelizing MASW will allow larger datasets to be analyzed in the field. In addition to parallelizing algorithms over CPUs, we can take advantage of other hardware: graphics cards. General purpose graphics processing units (GPUs), can be programmed to perform scientific computing simulations with significantly lower electricity cost and time per unit of computation. The majority of engineering workstations in offices today have a graphics card, even many gaming laptops that could be brought into the field, so a typical engineer could benefit from MASW software running on a GPU.

In this paper, we propose new algorithms to accelerate MASW imaging using MPI and GPUs, and introduce MASWAccelerated, a new open-source software package implementing these algorithms. In section 2, we give an overview of the serial algorithm for MASW, the MPI parallel algorithm, and the GPU accelerated algorithm. In section 3, we report on performance tests to show the speedup compared to MASWaves, an existing Matlab code \cite{Olafsdottir2018}, to show the scalability of the MPI algorithm, and to show speedups when utilizing a uniform-velocity dispersion curve and a more typical variable velocity dispersion curve. Further, we describe performance optimizations utilizing the sparsity structure of stiffness matrices, strategies to tackle load imbalancing which account for typical dispersion curve trends, and improved GPU performance when repeatedly evaluating the kernel as part of an optimization procedure. With these optimizations, the serial C implementation obtains a 2x speedup over MASWaves, the MPI implementation shows near perfect strong and weak scalability with uniform data, a modified partition of the problem for more realistic data shows near-perfect strong scaling up to 8 processes, and the GPU implementation obtains a 3.2x speedup over the serial C implementation. These speedups will enable engineers to perform MASW significantly faster, potentially even in the field.

\section{Overview of Algorithms}
We review the standard serial algorithm for MASW modeling from \citep{Olafsdottir2018}, and note optimizations we have made. We propose new algorithms that parallelize MASW modeling using MPI, and that perform MASW modeling on GPUs.

\subsection{Serial Implementation}

Before parallelizing the algorithm for MASW, we  need to implement a C serial version of the code which can then be modified with MPI and CUDA. We used MASWaves, an existing implementation written in MATLAB as our reference to compare against \cite{Olafsdottir2018}. Initially our C serial implementation was a simple port of MASWaves to C, but we later made a number of changes to the algorithm to improve its efficiency.

The model evaluation process in MASW consists of two main algorithms. First, the model parameter inputs are used to compute the most likely velocities for each wavelength in the experimentally derived dispersion curve. This is done in the function \verb|MASWA_theoretical_dispersion_curve|, illustrated in Algorithm~\ref{alg:theoDisp}. The model parameter inputs ($M$) are the number of finite thickness layers, the thickness and density of each layer, and the compressional (P) and shear (S) wave velocities through each layer.


\begin{algorithm}
\caption{MASW Theoretical Dispersion Curve}
\begin{algorithmic}[1]
\State $M$ = \texttt{Guessed model parameters}
\State $W$ = \texttt{Wavelength values from experimental dispersion curve}
\State $C_t$ = \texttt{Velocity corresponding to each wavelength given} $M$
\State $V$ = \texttt{Range of potential velocity values to test}
\State $d_{old}$ = \texttt{Determinant value for previous entry in $V$}
\State $d_{new}$ = \texttt{Determinant value for current entry in $V$}
\For{$w$ \texttt{in} $W$}
\State $d_{old}$ = \texttt{stiffness\_matrix}($M$, $V[0]$)
\State $d_{new}$ = \texttt{stiffness\_matrix}($M$, $V[1]$)
\State $n = 1$
\While{\texttt{sign}($d_{old}$) == \texttt{sign}($d_{new}$)}
\State $n = n+1$
\State $d_{old}$ = $d_{new}$
\State $d_{new}$ = \texttt{stiffness\_matrix}($M$, $V[n]$)
\EndWhile
\State $C_t[w]$ = $V[n]$
\EndFor
\end{algorithmic}\label{alg:theoDisp}
\end{algorithm}

MASWaves utilizes the stiffness matrix method (Kausel, 1981) to compute the theoretical dispersion curve. Given a wavelength, sparse stiffness matrices are computed for each test velocity (in $V$) in increasing order, until one has a determinant ($d_{new}$) with a different sign than its predecessor. This test velocity ($V[n]$) is then stored as the theoretical velocity corresponding to that particular wavelength, i.e. it is stored as $C_t[w]$. The process is repeated for each wavelength in the experimental dispersion curve to generate a theoretical dispersion curve of Rayleigh wave velocities and their wavelengths.

A stiffness matrix may be generated for each wavelength and each test velocity. Realistically a dispersion curve may have up to 100 wavelengths and 1,000 test velocities, requiring a total of 100,000 stiffness matrices to be generated (in practice this number is often lower, since the ideal test velocity is usually found before all velocities are checked for a given wavelength). These stiffness matrices also have a sparse banded structure, which is not utilized by MASWaves but is in our algorithms.

The algorithm also requires the determinants for each of these matrices to be computed. The stiffness matrices are of size $2(N+1) \times 2(N+1)$, where $N$ is the number of finite-thickness layers in $M$. Since they have a symmetric banded structure, computing the entries of a stiffness matrix and its determinant requires $O(N)$ operations.

\begin{algorithm}
\caption{MASW Misfit}
\begin{algorithmic}[2]
\State $C_{t}$ = \texttt{Velocity corresponding to each wavelength given} $M$
\State $C_{e}$ = \texttt{Experimentally derived velocities}
\State $e$ = \texttt{Relative errors}
\State $l$ = \texttt{Length of $W$, $C_t$, and $C_e$}
\State $m$ = the average misfit
\For{$i = 1$ to $l$}
\State e = e + $\frac{|C_t[i]-C_e[i]|}{C_e[i]}$
\EndFor
\State $m$ = $\frac{e}{l}$
\end{algorithmic}
\end{algorithm}

Once the theoretical dispersion curve $C_t$ is computed, the second part of the algorithm compares its velocities to the velocities experimentally derived from the data, labelled $C_e$. The average relative error is labelled the misfit, and indicates how accurate $M$ serves as a model of the ground structure. Essentially the misfit behaves as a loss function for MASW.

The model parameters $M$, along with the given test velocity and wavelength, are used to compute each stiffness matrix. If the theoretical velocity for each wavelength is close to the experimental velocity (i.e. the misfit is small), then the model parameters are more likely to be a good approximation of the ground truth.

One downside to MASW is that it does not have any form of backpropagation to accompany its loss function. Therefore the only way to minimize the misfit is to compute theoretical dispersion curves for an exhaustive quantity of plausible model parameters, each of which can require forming and finding the determinants of up to 100,000 small matrices. Since each of these curves and their misfits can be computed independently, this algorithm can benefit from parallelization, both between test models and within models.

\subsection{MPI Parallelism}

\begin{figure}[h]
\includegraphics[width=340pt]{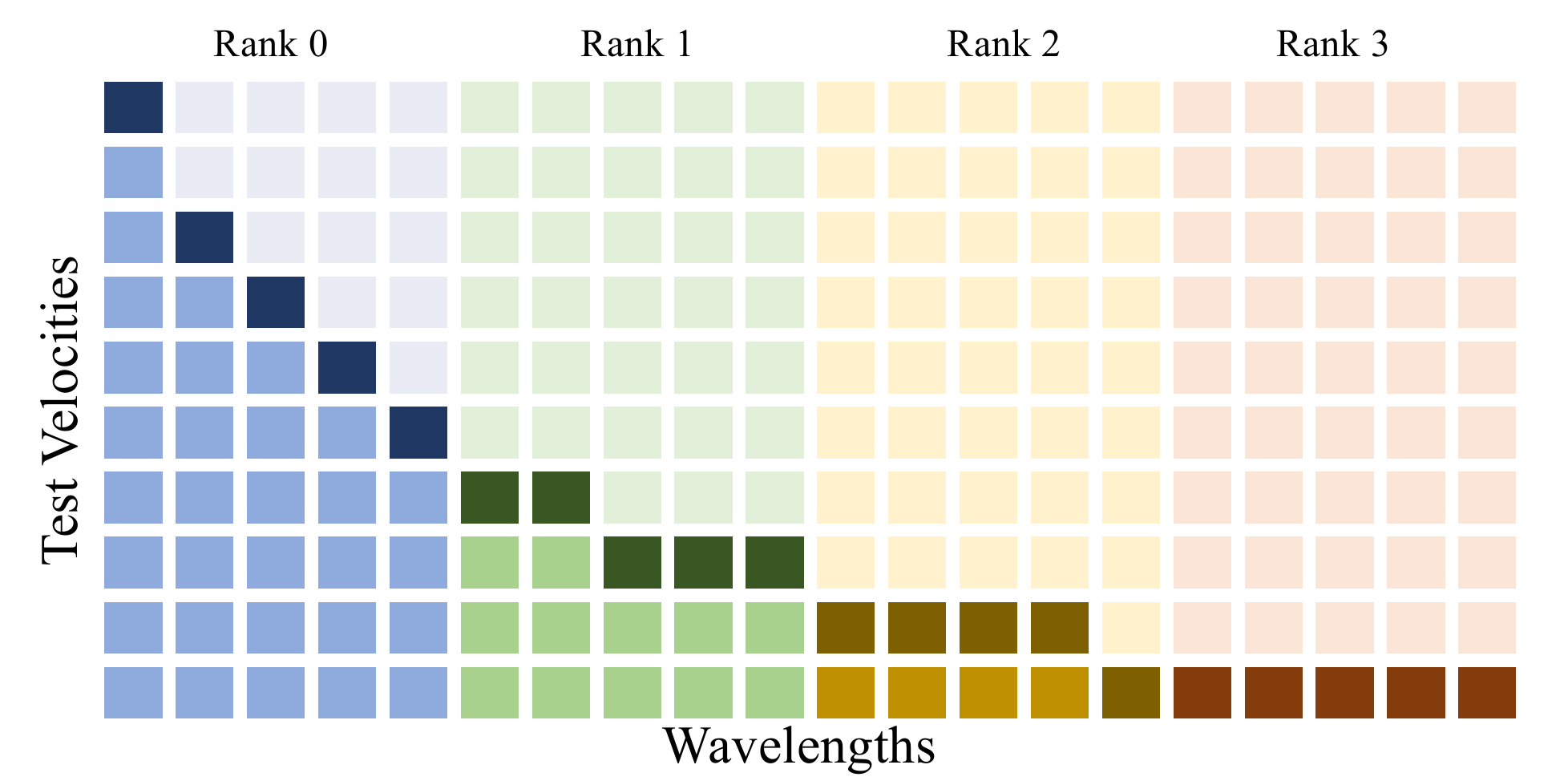}
\caption{Visualization of original MPI partition. Darker colors correspond to determinants that were computed, with the darkest highlighting the theoretical curve. Each rank has its own color scheme. Note the number of computed determinants varies heavily for each rank.}
\label{fig:fig0}
\end{figure}

It is possible to visualize MASW as a mesh computation, although it is not a literal ground mesh. Stiffness matrices must be computed with several different values of wavelengths and test velocities and the same values of model parameters. If the wavelength and velocity values are thought of as the x and y axis, then \verb|MASW_Theoretical_dispersion_curve| can be viewed as computing data points along a two dimensional ``grid". It is reasonable to partition this grid of computations into multiple processes in an MPI implementation.


There are two obvious ways to partition this grid. The stiffness matrix method heavily utilizes the test velocities in computing the individual matrices. Therefore, one could partition the grid along the velocity axis: given $S$ processes and $V$ test velocities, compute stiffness matrix determinants for all wavelengths and the first $\frac{V}{S}$ velocities in process 0, then the next $\frac{V}{S}$ velocities in process 1, and so on. This approach would allow components of the stiffness matrix dependent on the test velocity to be pre-computed, reducing the number of repetitive computations.

Partitioning along the velocity axis would have significant drawbacks, however. Once the determinants are computed, a linear search must be performed along each wavelength to find the first sign change. Since the determinants for each wavelength are split along multiple processes, a significant amount of communication would be required between each process to find the first sign change. In addition, determinants for higher test velocities might not even need to be computed if the first sign change (and thus correct test velocity) has already been found. Thus this approach may lead to several unnecessary computations.

Because of these difficulties, the method used was partitioning along the wavelength axis. The stiffness matrix computations do not feature the wavelength input as much as the test velocity, so there is little potential to pre-compute components of the matrices for each wavelength. But finding the ideal test velocity for a given wavelength has no dependence on other wavelength values, so no communication between processes is required for the dispersion curve. In addition, once the ideal test velocity is found for a particular wavelength, the process can begin computing matrix determinants for the next wavelength. This preserves the serial implementation's advantage of avoiding unnecessary matrix and determinant computations. Partitioning along the wavelengths also makes computing the misfit parallel as well, with only one reduction operation required to combine the errors from each process.

\begin{figure}[h]
\includegraphics[width=340pt]{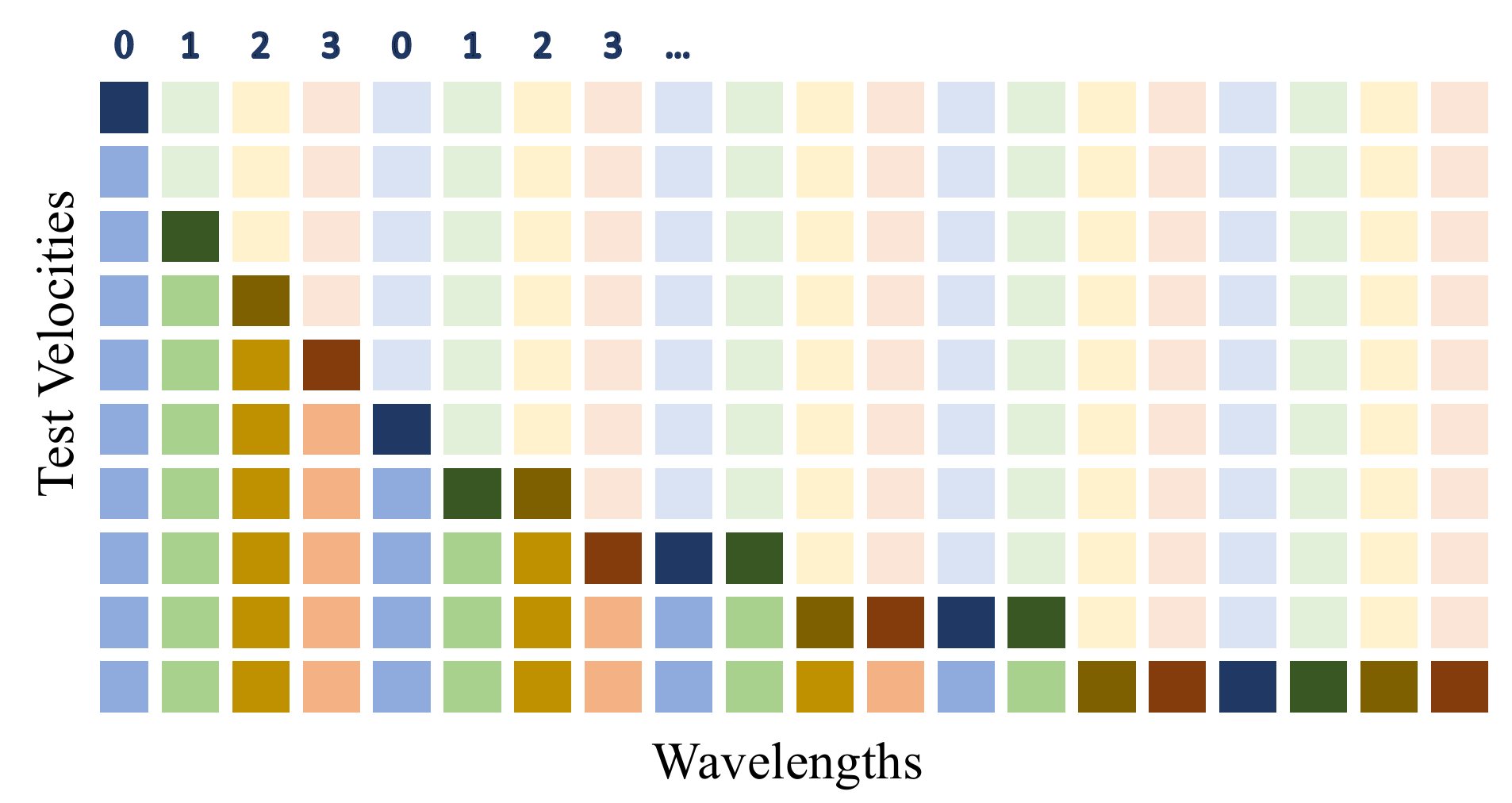}
\caption{Visualization of modular MPI partition, with the same color rules in place. In this case the computed determinants are balanced more evenly.}
\label{fig:fig-1}
\end{figure}

Since the algorithm does not require communication between different wavelengths of the dispersion curve, there is freedom to choose exactly how to partition the problem along the wavelength axis. The original approach was to assign wavelengths contiguously: given $s$ processes and $W$ wavelengths, assign wavelengths $0,1,\ldots,\frac{W}{s}$ to process 0, wavelengths $\frac{W}{s}+1,\ldots,\frac{2W}{s}$ to process 1, and so on. We determined a more efficient approach was to partition wavelengths in a modular pattern: assign every wavelength equivalent to 0 mod $s$ to rank 0, every wavelength equivalent to 1 mod $s$ to rank 1, and so on.

\subsection{GPU Acceleration}

\begin{figure}[h]
\begin{center}
\includegraphics[width=240pt]{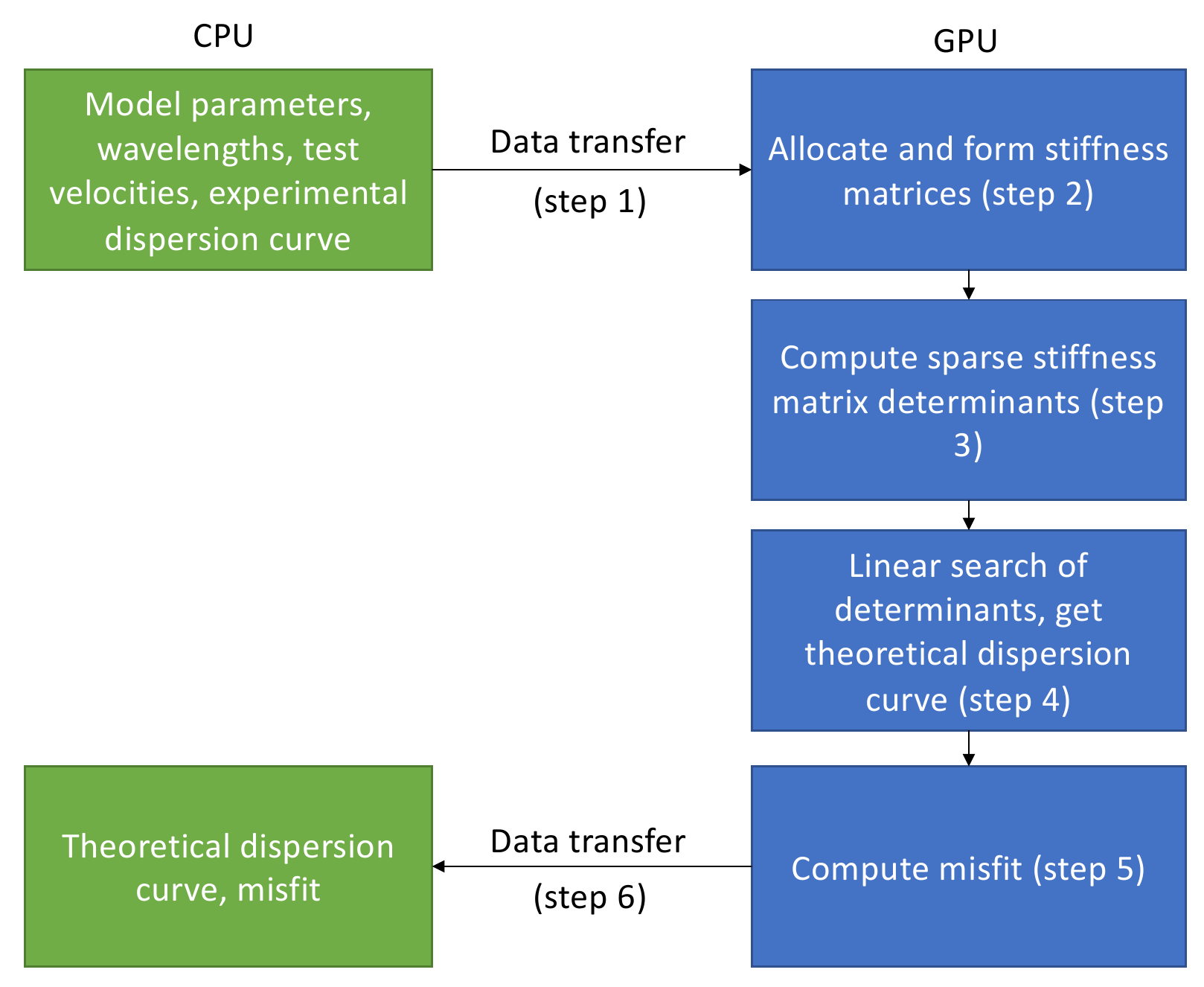}
\caption{General outline of GPU implementation.}
\label{fig:fig-2}
\end{center}
\end{figure}

Many workstations used by geotechnical engineers contain CPUs with multiple cores as well as graphics cards. Thus, in addition to writing MASW for MPI, we have also implemented the MASW algorithm for graphics processing units (GPUs). MASW requires calculating many determinants of sparse matrices, a problem which has not previously been adapted to GPUs.

There are five main steps to the GPU implementation that are distinct from the serial and MPI versions of MASW.

First, the initial input values to the algorithm - wavelengths, test velocities, experimental dispersion curve, and the ground model parameters - are transferred from the CPU ``host" to the GPU ``device" as shown in step 1 of Figure~\ref{fig:fig-2}. Copying memory between the host and device is costly, but the volume of data copied here is relatively small (the test velocity array may have $\approx 1000$ entries, and the other inputs are scalars or much shorter arrays), so this transfer is not problematic.

Once the inputs are on the device, the stiffness matrices are allocated as global device memory and their entries are filled in based on the model parameters $M$, shown in step 2 of Figure~\ref{fig:fig-2}. Forming the stiffness matrices is one of the most costly steps in MASW, so effective parallelization is critical. Since a GPU contains thousands of cores (as opposed to the dozens that may be available on one or more CPUs), it is feasible to partition the problem along both the wavelength and velocity axis dimensions of the grid. In fact, it is reasonable to compute all the stiffness matrices concurrently since they are mutually independent. Using CUDA, each thread is assigned to fill in the entries for one stiffness matrix.

Computing values for a stiffness matrix is $O(N)$ as described in the serial implementation, where $N$ (the number of finite thickness layers in $M$) is typically small ($N \leq 10$). Entries in the stiffness matrices are incrementally increased multiple times, so trying to compute a stiffness matrix across multiple threads can lead to race conditions. Thus assigning a single thread to each stiffness matrix is reasonable.

Once the stiffness matrices are formed, Gaussian elimination is performed so their determinants can be easily computed, shown in step 3 of Figure~\ref{fig:fig-2}. This is roughly equal to forming the stiffness matrices in terms of time cost. Initially, we used the function \verb|cublasZgetrfBatched()| to perform LU factorizations on the stiffness matrices. This function took over 50\% of the runtime of the GPU implementation, most likely because it did not take advantage of the stiffness matrices' banded structure and was therefore $O(N^3)$ for each factorization. Because of this, we replaced it with a kernel that assigned one thread to each stiffness matrix to perform a banded Gaussian elimination. This function was roughly ten times faster than \verb|cublasZgetrfBatched()|, and approximately equal in runtime to the kernel that formed the stiffness matrices.

The next step is to find the first test velocity whose corresponding determinant is the opposite sign of its predecessor for each wavelength, which is step 4 in Figure~\ref{fig:fig-2}. Normally this would require a linear search along all the test velocities for each wavelength - a serial $O(N)$ process that does not lend well to GPU architecture. However, there is still some potential to partition the problem along the device cores. The search kernel breaks up the stiffness matrices along its blocks by wavelength, and along threads within each block by test velocity. This is illustrated in Figure~\ref{fig:fig-3}:

\begin{figure}[h]
\begin{center}
\includegraphics[width=300pt]{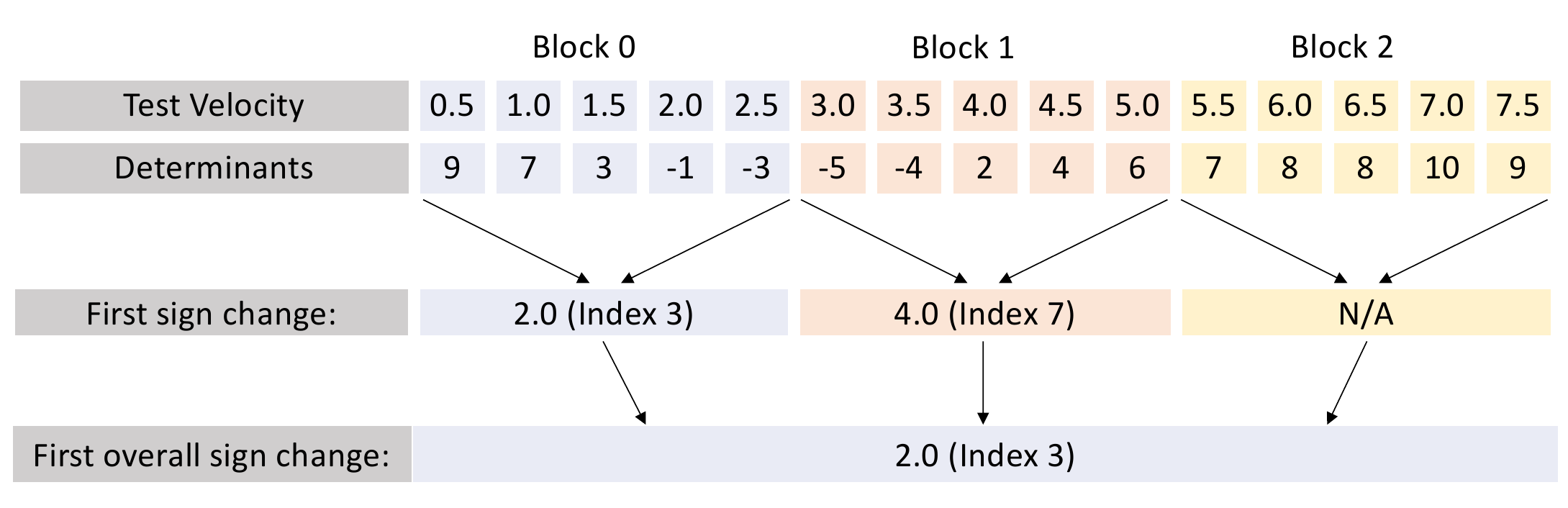}
\caption{Outline of GPU determinant search along one wavelength. The first sign change within each block is found, then a search over the blocs finds the first overall sign change.}
\label{fig:fig-3}
\end{center}
\end{figure}

Each wavelength will be paired with multiple blocks, since the block size is assumed to be 256 (for compatibility with older GPUs) and often MASW is run with $\geq 500$ test velocities. Within each block,  thread $i$ computes the determinant of its respective stiffness matrix by multiplying the diagonal entries, then compares it to the determinant of thread $i+1$. It then stores the result of the sign comparison into shared memory.

Once this is complete for all threads, the first thread of each block then performs a linear search for the first sign change in its shared memory block, and stores this as the index of the first sign change within that range of test velocities. These results are placed in the matrix $\in \mathbb{R}^{\ell \times b}$, where $\ell$ is the number of wavelengths and $b$ is the number of blocks assigned to each wavelength. The next kernel then iterates over $S$ to find the first recorded sign change for each wavelength, which is a small linear search over approximately 4 entries. The test velocity whose stiffness matrix determinant produced the first sign change is then labelled as the velocity corresponding to that wavelength in the theoretical dispersion curve.

The final steps are to average the errors between the theoretical dispersion curve and experimental curve to get the overall misfit, and to send the theoretical dispersion curve and misfit to the host, which are steps 5 and 6 of Figure~\ref{fig:fig-2}. The former is effectively equivalent to a vector dot product (in terms of number of FLOPs), which is straightforward to implement in CUDA. The latter is simply a CUDA memory copy involving a vector of length $\leq 100$ and a scalar. Steps 1 and 6 require only a small amount of data to be transferred between the host and device, while the stiffness matrices, which take up much more memory, are allocated and freed exclusively on the device.

\section{Test Results and Performance Optimizations}

We developed a variety of tests for correctness of the code (unit tests and end-to-end tests), as well as performance and scalability tests to understand code efficiency. 

We used two main test cases to evaluate MASWAccelerated's performance. A synthetic dispersion curve of identical wavelength values was used to provide a ``uniform" dataset. This is useful because of the design of the serial and MPI algorithms - since every test velocity is evaluated in increasing order, varying wavelengths with different theoretical velocities will have different runtimes. Thus a uniform test allows us to observe other factors that may affect how MASWAccelerated scales with more data. The second test case was a more realistic ``variable" dataset, with decreasing wavelengths, matching a typical dispersion curve. We present these test cases in the code as \verb|testScaling| and \verb|testProcess|, respectively. Both of these functions also have wrappers (\verb|testScaling_full| and \verb|testProcess_full|) that enable test cases to be run multiple times in a loop.

\subsection{Serial Tests}

\begin{figure}[h]
\includegraphics[width=340pt]{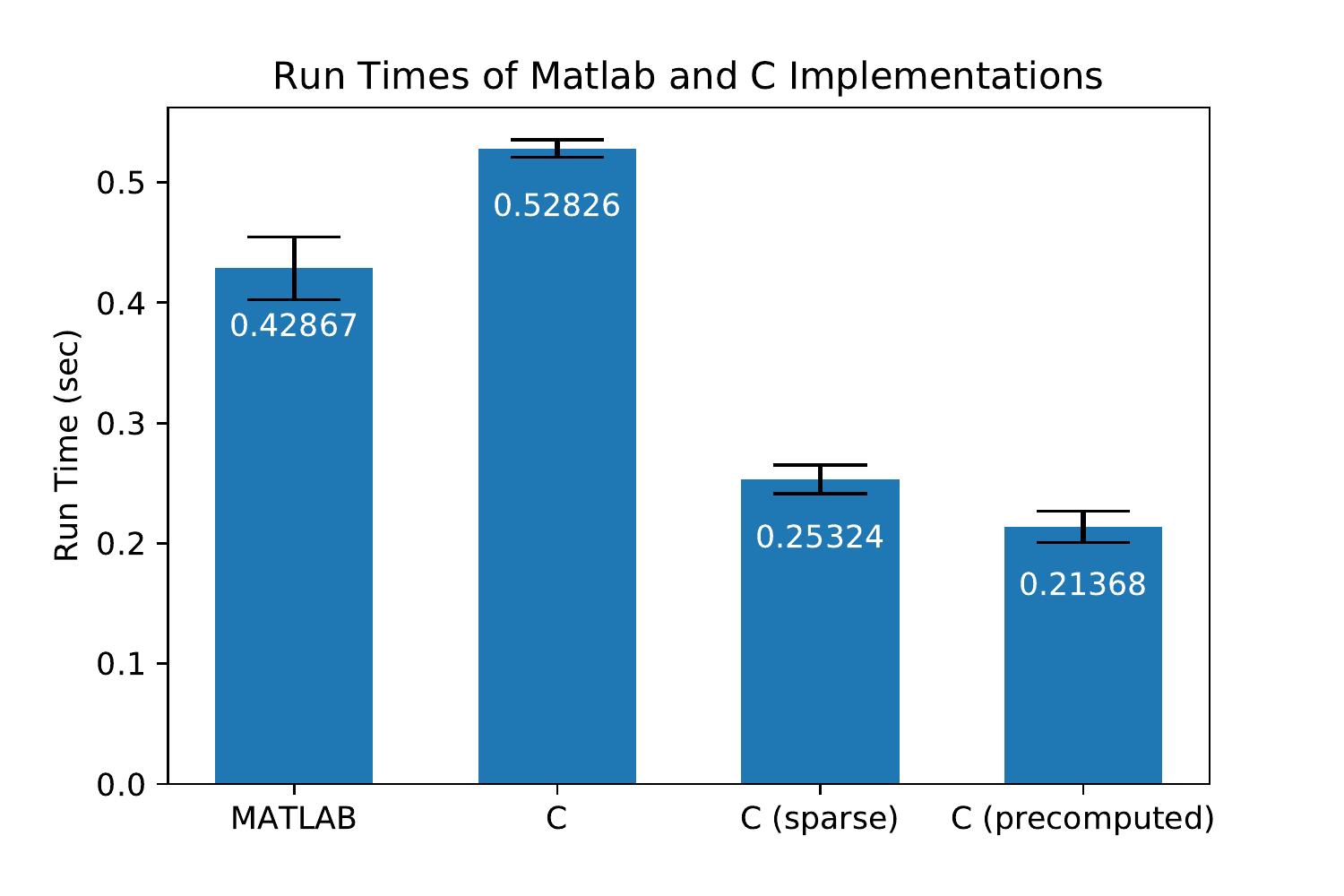}
\caption{Comparison of MATLAB and C on a variable dataset.}
\label{fig:fig1}
\end{figure}

The primary purpose of the serial C implementation is to enable usage of MPI and CUDA to parallelize MASW. However, it is still important to make sure the serial C version is correct, and it is useful to compare its speed to the original MATLAB version, MASWaves. We timed the speed of MASWaves on our realistic dataset, and compared it to MASWAccelerated's speed on the variable dataset when run in serial. Both of these tests were run ten times on a laptop with a 3.1 GHz Intel Core i7 CPU, and the mean results are shown in Figure~\ref{fig:fig1}, along with errors denoting one sample standard deviation.

Initially the serial C implementation, while mathematically correct, was slower than MASWaves, which is not ideal. This is likely because MASWaves made use of MATLAB's vectorized operations to compute stiffness matrix determinants, while the initial Gaussian elimination algorithm written for MASWAccelerated was not vectorized.

The stiffness matrices formed by MASW are always sparse, and moreover have a banded heptadiagonal structure (nonzero entries only on the main diagonal and the three above and below it). We used this fact to improve the Gaussian elimination algorithm to be only $O(N)$ instead of $O(N^3)$, which increased the algorithm's speed by 2.0 times, as shown in the C (sparse) column. We also noticed components of the entries in the stiffness matrices could be pre-computed before iterating over the matrix entries in a loop, reducing the number of arithmetic operations required and increasing the algorithm speed a further 18\% as shown in the time for C (precomputed). These improvements enable the serial C implementation to be over twice as fast as MASWaves without any parallelization or vectorization calls. The precomputed serial C version is used as the basis for the MPI algorithm, and for performance comparison with the GPU algorithm.

It is worth noting the runtime for the first instance of MASWaves was significantly slower than subsequent runs, causing the increased error. This may be due to some type of caching effect in MATLAB, which is not present in C.

\subsection{MPI Tests}

As the MPI implementation with one process is virtually identical to the serial implementation, the main purpose of testing is to evaluate how it scales with more processes and larger datasets. For this purpose we ran a few strong and weak scaling studies.

\begin{figure}[h]
\includegraphics[width=340pt]{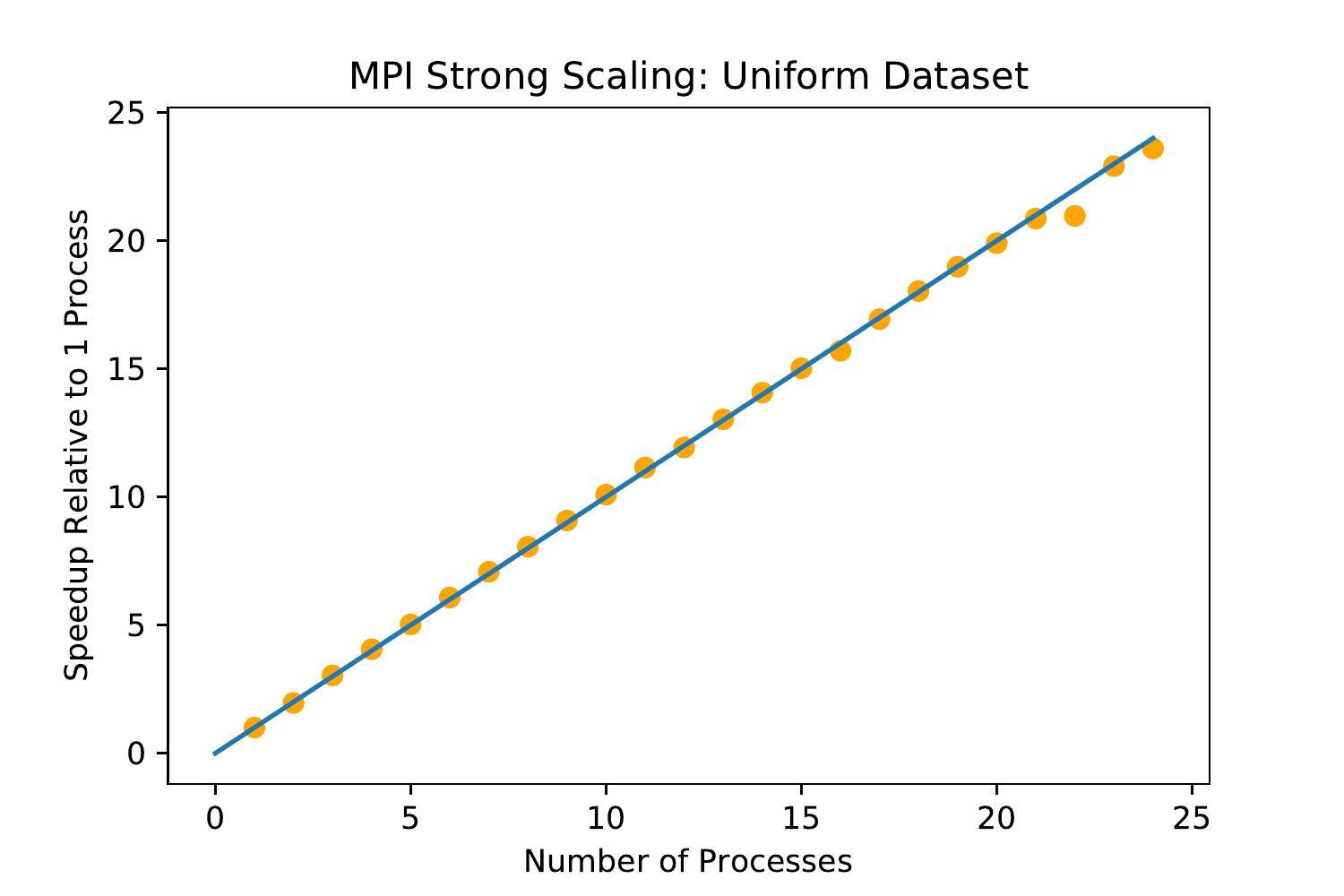}
\caption{MPI strong scaling on uniform data.}
\label{fig:fig3}
\end{figure}

For the strong scaling study, we first tested the algorithm on the uniform dataset with 1000 wavelengths, allowing for distinctions in runtime to be more noticeable. The results are shown in Figure~\ref{fig:fig3}, which shows the average runtime for each process count on ten runs.

This test was run on the NewRiver computing cluster at Virginia Tech, using two Haswell E5-2680v3 2.5GHz processors with 12 cores each. The algorithm scaled almost exactly linearly with more cores (except at 22 cores, which may be due to some error), which is expected given the minimal amount of communication required between processes and the highly parallel method used to partition MASW.

In practice, dispersion curves usually have a couple features that make the MPI algorithm scale less than linearly, as seen in the strong scaling test for the uniform dataset.

One problem is that dispersion curves do not have identical wavelength values throughout, but rather varying wavelength values which correspond to different velocities. Since MASW must evaluate test velocities in increasing order to identify the determinant sign change, entries with larger velocities will require more stiffness matrix determinants to be computed, resulting in a longer run time. Because of this, the MPI algorithm is prone to load imbalancing on realistic data, even though the entries of the dispersion curve are partitioned evenly.

\begin{figure}[h]
\includegraphics[width=340pt]{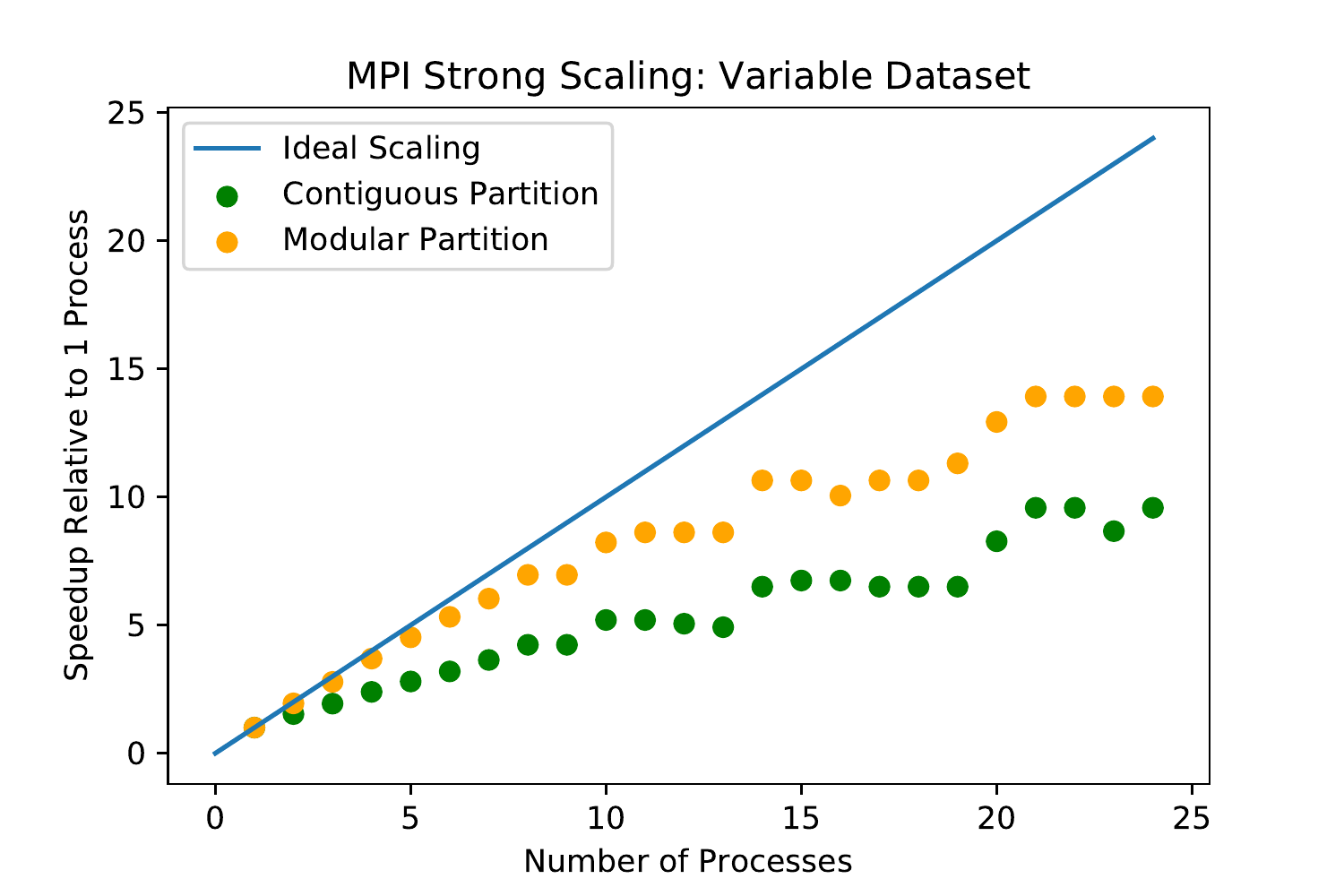}
\caption{MPI strong scaling on variable data, using different partitions. The variable data highlights the two features that can make MASW problematic for MPI: large variations in wavelength values for the dispersion curve, and short dispersion curve length.}
\label{fig:MPIStrongVariable}
\end{figure}

For near-surface imaging, most dispersion curves are decreasing in both wavelength and velocity, and usually resemble a continuous curve. A naive contiguous partition of the dispersion curve will place entries with similar wavelengths and velocities on the same process, thus resulting in a few processes receiving all of the high-velocity entries, exacerbating the load imbalance. The modular partition of the dispersion curve mitigates this problem, and improves the MPI algorithm's strong scaling as a result. The comparison of speedup for these partitions on the variable dataset can be seen in Figure~\ref{fig:MPIStrongVariable}. A line has been added to compare both to ideal linear scaling.

This test was run on the same hardware as the uniform strong scaling, and again the average of ten runs per size was used. Although still not linear, the modular partition scales significantly better than the original contiguous partition: with 3 processes it has a speedup of nearly 2.8 compared to 1.9 for the naive approach, and with 8 processes it has a speedup of nearly 7.0 compared to only 4.2. Overall, the MPI algorithm with a modular partition will have near-linear scaling for most datasets with relatively few processes.

The reduction in speedup with more processes is likely due to another problem - most dispersion curves have a relatively short length (the variable dispersion curve has 40 entries, which is fairly typical). When the size of the partition is small, approximately 10 or less, each additional process reduces the number of dispersion curve entries computed for all of the other processes. For example, at size 3 each process is computing velocities for 13 or 14 wavelengths, while at size 4 each process is computing velocities for only 10 wavelengths. This is a significant reduction in workload and results in major speedup as seen in Figure~\ref{fig:MPIStrongVariable}. But at larger sizes there is not a reduction in workload for every process. For example, at size 20 each process is computing 2 entries, while at size 24 16 processes are still computing 2 entries and the last 8 are computing 1. Since many processes have no reduction in workload, the overall runtime of the algorithm is not reduced. Therefore, the MPI partition scales near linearly with relatively few processes (depending on the size of the dispersion curve), but experiences diminishing returns with more processes.

\begin{figure}[h]
\includegraphics[width=340pt]{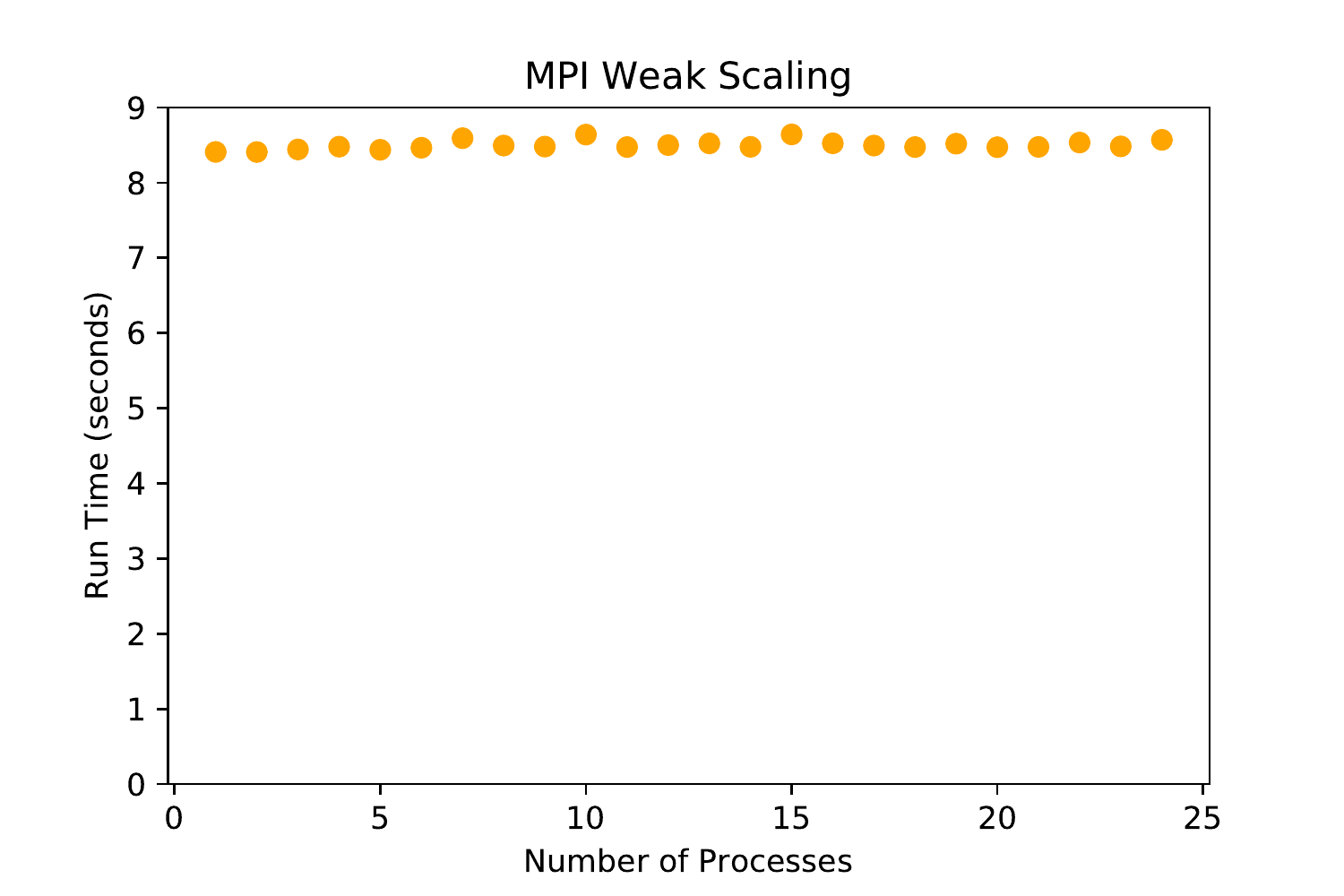}
\caption{MPI weak scaling on uniform data, with a dispersion curve of length 1000 $\times$ number of processes.}
\label{fig:MPIWeak}
\end{figure}

A weak scaling study was also performed on the NewRiver cluster using the uniform dataset, again with the same hardware and taking the average of ten runs per size. The size of the dispersion curve was 1000 $\times$ the number of processes. As seen in Figure~\ref{fig:MPIWeak}, the MPI implementation scaled efficiently by this measure, having no significant increase in time with more data and processes. There are a few sizes with marginally higher runtimes - 7, 10, and 15 - but these are minor and likely due to external factors. Like the strong scaling study with uniform data, this test highlights the minimal communication requirements for the MPI implementation.

\subsection{GPU Tests}

\begin{figure}[h]
\includegraphics[width=340pt]{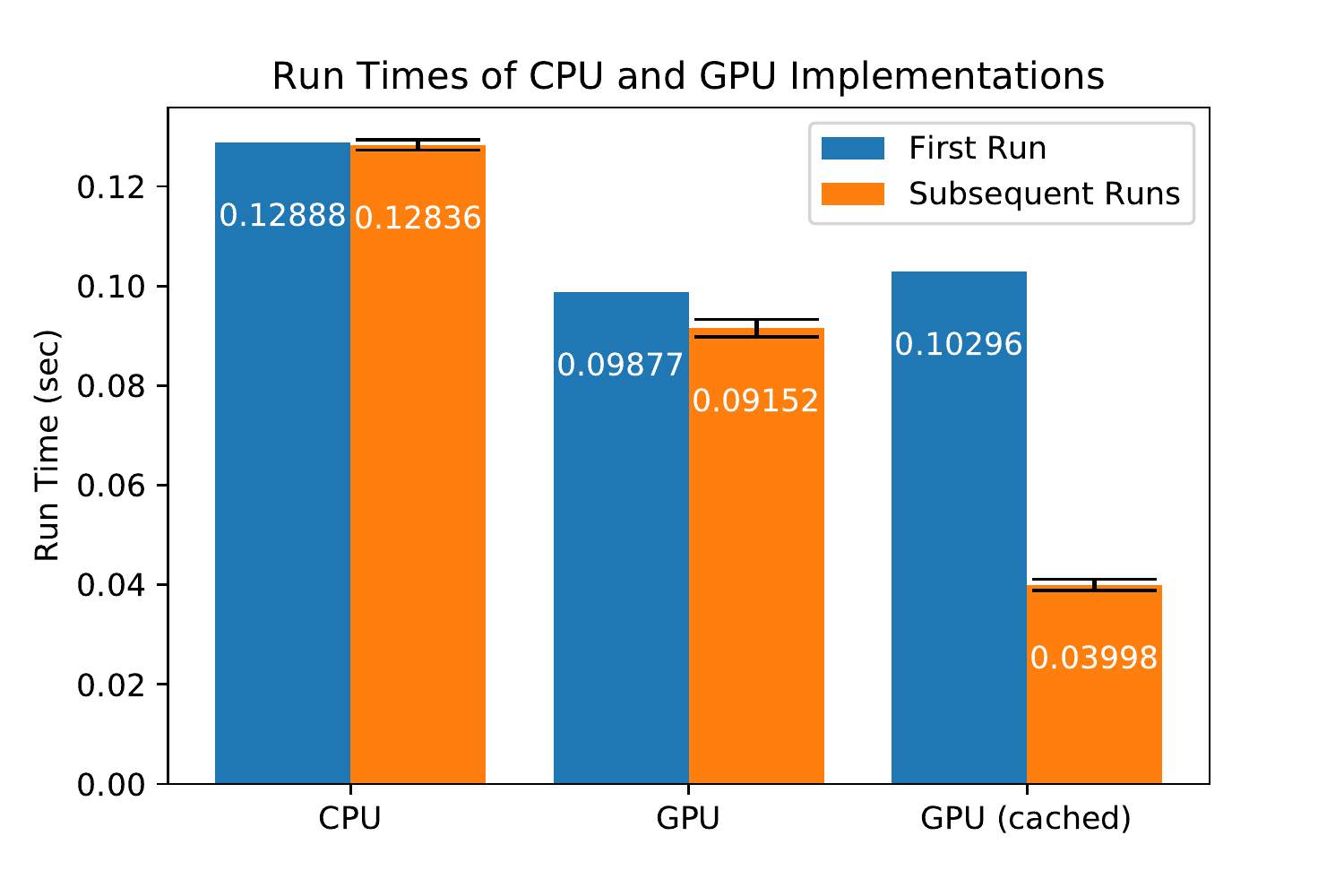}
\caption{Comparison of CPU and GPU on the variable dataset.}
\label{fig:fig6}
\end{figure}

First we compared the CPU and GPU implementations on the variable dataset. This was done eleven times in the wrapper loop \verb|testProcess_full| using the MPI implementation (with one process), then eleven times both in a loop and separately with the CUDA implementation. The resultant run times are shown in Figure~\ref{fig:fig6}. Note the blue bars are the first run for each method, while the orange bars are the means of subsequent runs with their sample standard deviation posted. The CPU algorithm is unchanged from the serial and MPI tests, but its runtime is different since it was run on a different machine.

All three of these implementations were run on the same desktop, using an Intel Xeon CPU E3-1271 v3 @ 3.60 GHz and an Nvidia Quadro K620 GPU. The first run is the run time of the first instance of \verb|MASW_inversion|, while the subsequent runs denote the mean runtime of all other instances. The error bars denote one standard deviation for the subsequent runs. The GPU implementation of \verb|MASW_inversion| is about 25\% faster than the CPU implementation when run once, but over 3.2 times faster when it is run multiple times in a for loop. This is likely because of just-in-time compilation, used by Nvidia to allow CUDA kernels to benefit from new device architectures. When a kernel is run multiple times within a function call, it only needs to be compiled for the first kernel run while it is ``cached" for subsequent runs. Since MASW inversion is often run multiple times with different test velocity models, it is reasonable to design MASWAccelerated to run inversions on multiple models to take advantage of this caching effect. It is worth noting that, while not as dramatic, there is still roughly 8\% speedup when running the inversion multiple times individually on the GPU, but no significant speedup for the CPU implementation.

\begin{figure}[h]
\includegraphics[width=340pt]{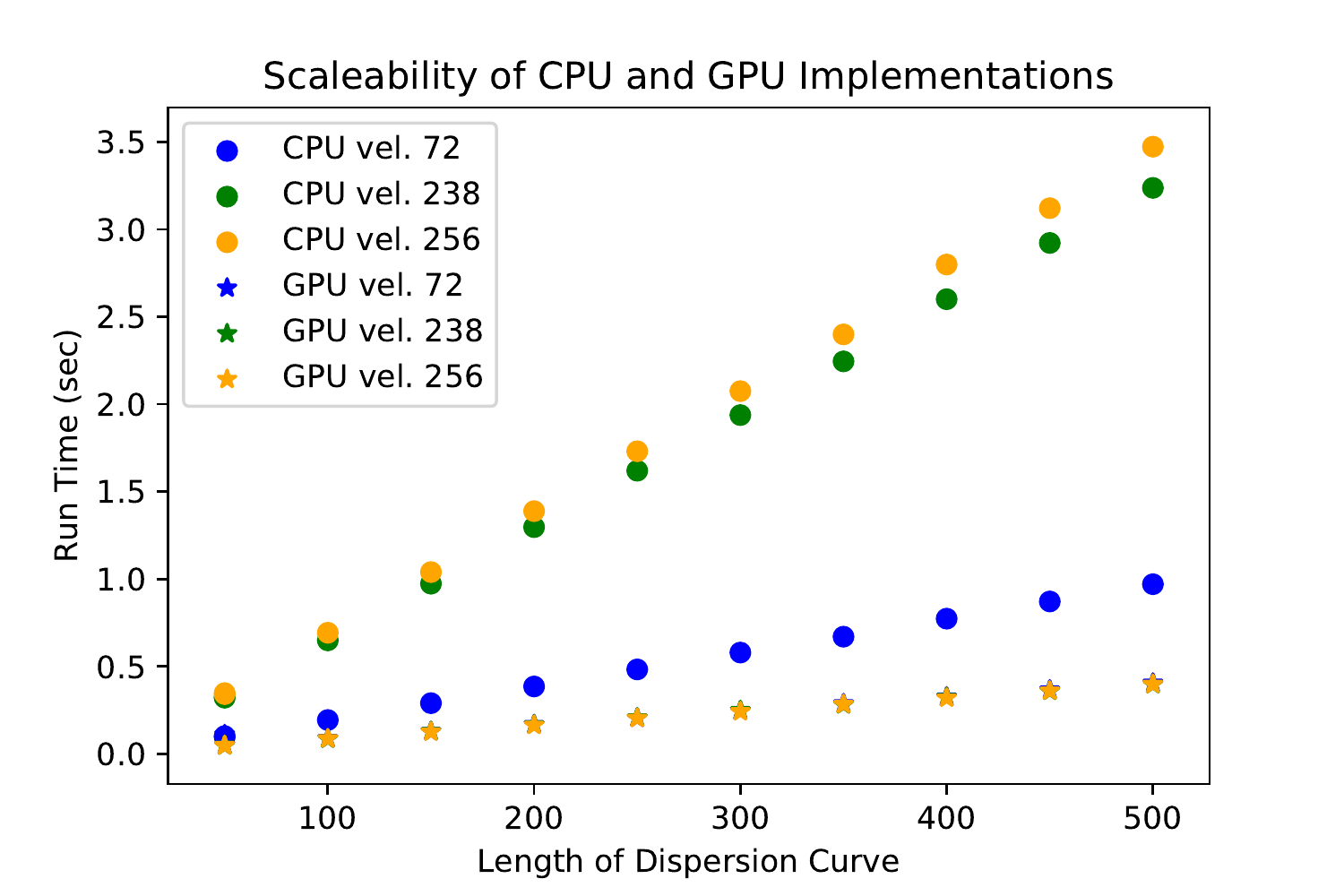}
\caption{Comparison of CPU and GPU on increasing uniform datasets.}
\label{fig:fig7}
\end{figure}

We next use the uniform dispersion curve to evaluate how the GPU implementation performs on progressively larger datasets. This test ran the MPI (one rank) and CUDA implementations of \verb|MASW_inversion| on dispersion curves of lengths 50 - 500. The wavelength values of these curves were designated to match test velocities of 72, 238, and 256 (each implementation and each dispersion length was run three times with three different wavelength values). The results are shown in Figure~\ref{fig:fig7}, run on the same desktop used for the previous test.

As expected, the CPU implementation scaled linearly with the length of the dispersion curve. The exact run time is highly dependent on the theoretical velocity values, as shown in the MPI scaling studies. The GPU implementation was significantly faster (since the caching effect was recognized in testing the variable data, we made use of it here), and had no dependence on the theoretical velocities since all stiffness matrices are computed regardless.

It is worth noting that too large a dispersion curve or too many test velocities can overload the GPU memory on the CUDA implementation, since all the stiffness matrices are formed concurrently in global memory and the dispersion curve and test velocities determine the number of stiffness matrices. The exact upper limit depends on the memory space of the GPU and the number of finite thickness layers in the model $M$ (which determines the size of the stiffness matrices), but typical problem sizes for MASW will not take up too much memory, even for older GPUs. We found dispersion curves larger than 500 typically caused memory problems for the Quadro K620 GPU, which has 2 GB of global memory. This is because each stiffness matrix has 196 entries, each of which is a CuDoubleComplex datatype that takes 16 bytes, so 500 wavelengths $\times$ 1000 test velocities $\times$ 3136 bytes $\approx 1.6$ GB, close to the memory limit of the GPU.

\section{Discussion and Conclusions}

There are two major priorities to improve the usability of MASWAccelerated. First, an I/O system would make it much easier to input model data and collect results from the inversion process. It would also eliminate the need to recompile the code each time new data is being inputted. The other priority is enabling command line arguments, such as the option to evaluate multiple input files (with separate data models) at once, or to run the suite of test functions. These two enhancements are necessary to enable the next step of MASWAccelerated being easily applied to a wide variety of use cases.

Another useful enhancement would be integrating the MPI and Cuda implementations of MASWAccelerated into one program. Theoretically, this is feasible, as each process in MPI is essentially MASW inversion carried out only on a portion of the dispersion curve, which can still be done by a GPU.  Solving this problem would enable further optimization of the algorithm to various hardware configurations.

In this paper, we have proposed two new algorithms for MASW: one using MPI parallelism, and one accelerating the code with Cuda for GPUs. We provide an open-source implementation called MASWAccelerated, along with test cases to verify code correctness and efficiency. Our comparisons show significant speedups over the existing MASWaves code, and we show several optimizations that take advantage of the problem structure (both typical trends in data and sparsity structure of stiffness matrices) to further improve code efficiency. These modifications will help engineers to perform rapid data analysis by taking advantage of all computer hardware available to them, ideally even performing analysis in the field as data are acquired.

\section{Acknowledgements}
We would like to thank the DOE Office of Geothermal Technologies for financial support through the STTR Phase I Award No. DE-SC0019630 which is a collaboration with Luna Innovations led by P.I. Derek Rountree. We thank Virginia Tech Advanced Research Computing for computing resources. 

This material is based upon work supported by the U.S. Department of Energy, Office of Science, SC-1 under Award Number DE-SC0019630. Disclaimer: This report was prepared as an account of work sponsored by an agency of the United States Government. Neither the United States Government nor any agency thereof, nor any of their employees, makes any warranty, express or implied, or assumes any legal liability or responsibility for the accuracy, completeness, or usefulness of any information, apparatus, product, or process disclosed, or represents that its use would not infringe privately owned rights. Reference herein to any specific commercial product, process, or service by trade name, trademark, manufacturer, or otherwise does not necessarily constitute or imply its endorsement, recommendation, or favoring by the United States Government or any agency thereof. The views and opinions of authors expressed herein do not necessarily state or reflect those of the United States Government or any agency thereof.

\section{Computer Code Availability}
The MASWAccelerated software, along with examples to produce the results in this paper, are publicly available at \url{https://github.com/jlk9/MASWA} under an MIT license. The code was primarily developed by Joseph Kump, who can be reached by email at josek97@vt.edu, or at the Virginia Tech Department of Mathematics, 225 Stanger St., Blacksburg, VA, 24060, ph: (540)231-6536. The MASWAccelerated code was first made publicly available in 2020 upon submission of this paper. Hardware and software requirements, as well as other code features, are available in the repository's README file.


\bibliography{mybibfile}

\end{document}